\documentclass{elsart}
\usepackage{psfig,latexsym,amssymb}

\begin{document}

\begin{frontmatter}

\title{Integrable open boundary conditions for the Bariev model
of three coupled XY spin chains}
\author[PortoAlegre]{A.~Foerster},
\author[Queensland]{M. D. ~ Gould},
\author[PortoAlegre]{X.-W.~Guan},
\author[RiodeJaneiro,Sunysb]{I.~Roditi},
\author[Queensland]{H.-Q~Zhou},
\address[PortoAlegre]{Instituto de Fisica da UFRGS,
                     Av.\ Bento Goncalves, 9500,\\
                     Porto Alegre, 91501-970, Brasil}
\address[Queensland]{
Department of Mathematics, \\ University of Queensland, Brisbane, Qld 4072, Australia}
\address[RiodeJaneiro]{Centro Brasileiro de Pesquisas Fisicas,\\Rua Dr. 
Xavier Sigaud 150, 22290-180, Rio de Janeiro- RJ, Brasil}
\address[Sunysb]{C N Yang Institute for Theoretical Physics, State
University of New York at Stony Brook, NY 11794-3840, USA}

\begin{abstract}
The integrable open-boundary conditions for the Bariev model of three
coupled one-dimensional $XY$ spin chains are studied in the framework
of the boundary quantum inverse scattering method.  Three kinds of 
diagonal boundary K-matrices leading to nine classes of possible
choices of boundary fields are found and the corresponding integrable
boundary terms are presented explicitly.  The boundary
Hamiltonian is solved by using the coordinate  Bethe ansatz
technique and the Bethe ansatz equations are derived.
\end{abstract}


\end{frontmatter}



\def\a{\alpha}
\def\b{\beta}
\def\d{\delta}
\def\e{\epsilon}
\def\g{\gamma}
\def\k{\kappa}
\def\l{\lambda}
\def\m{\mu}  
\def\o{\omega}
\def\t{\theta}
\def\s{\sigma}
\def\D{\Delta}
\def\L{\Lambda}


\def\beq{\begin{equation}}
\def\eeq{\end{equation}}
\def\bea{\begin{eqnarray}}
\def\eea{\end{eqnarray}}
\def\ba{\begin{array}}
\def\ea{\end{array}}
\def\no{\nonumber}
\def\le{\langle}
\def\re{\rangle}
\def\lt{\left}
\def\rt{\right}

\newcommand{\sect}[1]{\setcounter{equation}{0}\section{#1}}
\renewcommand{\theequation}{\thesection.\arabic{equation}}
\newcommand{\reff}[1]{eq.~(\ref{#1})}

\vskip.3in

\sect{Introduction\label{int}}

Since the discovery of high temperature superconductivity in cuprates
\cite{BM86}, a tremendous effort has been made to uncover a
theoretical framework capable of explaining this amazing phenomenon.
It is a general belief that the properties of strongly correlated
electron systems showing a non-Fermi liquid behaviour are closely
related to those materials showing high $T_c$ superconductivity.  This
has caused an increasing interest in strongly
correlated electron models \cite{[1],[2],[3],[4]}. Integrable models
that have been widely studied include the 1D Hubbard model solved by 
Lieb and Wu \cite{LW} and
the supersymmetric t-J model \cite{azr,tj}. Another model with special
features relevant for high $T-c$ superconductivity is the 1D Bariev
\cite{Bar1,Bar2},  as it exhibits the existence of
hole pairs of Cooper type.

On the other hand, important progress in the realm of completely
integrable
systems is the generalization of the usual quantum inverse scattering
method (QISM) \cite{qism1,qism3,qism4} to incorporate 
open boundary conditions that preserve integrability
\cite{op1,op2,op3,op4}. The presence of the integrable boundaries
leading to a pure back-scattering on each end of a quantum chain 
results in rich physical phenomena\cite{Frah,Asak,Wad1,be1,be2,be3}.
It has been clarified that 1D quantum systems with boundary fields are
closely related to impurity problems.  Also due to their connection to the
Kondo problem and boundary conformal field theory in low-dimensional
quantum many-body systems, the integrable boundaries could be a useful
non-perturbative way to investigate the impurity effects in the
condensed matter physics. The prototypical 1D Hubbard model and t-J
model have been thoroughly investigated  with boundary impurities and
boundary fields \cite{Frah,Asak,Wad1,be1,be2,be3,op5,op6}. Subsequently, other
strongly correlated electron systems, such as the 1D Bariev model of
two coupled XY spin chains have been partially studied
and the reflections equations (RE) solved\cite{bariev}. In
that case, a new class of boundary reflection K-matrices for the model
lead to pure magnetic boundary fields in the Hamiltonian which may have a
feasible realization by applying boundary external fields in
experiments on quantum wires. An interesting aspect here is that after
Jordan-Wigner transformation, the coupled XY spin chains can be
presented as correlated electron systems where the hopping terms
depend on the occupation numbers at sublattices. This is thought to
be useful in studying conductivity properties in such non-fermi
liquids. Along this line, the integrability and boundary conditions
for a Bariev model of three coupled XY chains has been studied in the
frame work of the QISM by Zhou and coworkers \cite{Zhou97,zhou3}
recently. However, we have good reasons to expect that the model
would permit other kinds of integrable boundary terms associated with
some new solutions to the RE. We revisit the model
seeking a complete understanding of the open boundary
conditions for the model.  As noticed in \cite{zhou3}
the quantum $R$-matrix of the model we study does not possess the
crossing-unitarity. This causes a violation of the isomorphism between
$K_+$- and $K_-$-matrix which satisfy two reflection equations (RE)
separately. The integrability at left and right
boundaries demand rather complicated RE which in turn expose
bulk symmetries on the left and right boundary terms. In order to maintain
the integrability of models where the $R$-matrix does not posses the
crossing-unitarity property some new objects have
to be introduced .  It is found that the three coupled XY chains permits three
kinds of K-matrices to each RE which lead to nine classes of
integrable boundary terms containing the ones for the two coupled XY
spin chain as special cases. These integrable boundary terms containing
different on-site Coulomb interactions and chemical potentials reveal
the symmetry of exchanging the sublattices. We derive the Hamiltonian
of the model with nine classes of boundary terms from the expansion of
the boundary transfer matrix around the zero spectral-parameter point
up to the fourth, third and second orders respectively.  Furthermore,
we solve the boundary model by means of the coordinate Bethe ansatz
method and derive the Bethe ansatz equations.

The paper is organized as follows. In section 2 we construct the
Bariev model of three coupled XY spin chains with nine classes of
boundary fields by means of the QISM adapted to special boundary
conditions. The basic quantities, e.g., the $R$-matrix, the matrices
$K_{\pm}$ defining the boundary terms, the monodromy matrices and the
transfer matrices are defined.  In particular, we solve two RE
separately and obtain three independent classes of solutions to 
each REs using the variable-separation method.
The relation between the transfer matrices and the Bariev Hamiltonians
with different boundary fields is established.  Section 3  is 
devoted to the solution of the models through coordinate Bethe ansatz
method. Section 4 presents our conclusions.

\sect{Boundary K-matrices for the coupled spin chains \label{Boun}}

We consider a spin chain model defined by the following Hamiltonian
\beq
H = \sum ^{L-1}_{j=1} H _{j,j+1}+B^{(m)}_1 +B^{(l)}_L,\label{hamiltonian}
\eeq
where $H_{j,j+1}$ denotes the bulk Hamiltonian density of
three $XY$ spin chains coupled to each other \cite{Zhou97}
\bea
H_{j,j+1} & = & \sum _{\alpha}\;(\sigma^+_{j(\alpha)}\sigma^-_{j+1(\alpha)}+
\sigma^-_{j(\alpha)}\sigma^+_{j+1(\alpha)})\\
& &\exp[\eta \sum _{\alpha' \neq \alpha }
\sigma ^+_{j+\t(\alpha'-\alpha)(\alpha')}
\sigma ^-_{j+\t(\alpha'-\alpha)(\alpha')}],\label{bulk-h}
\eea
where $\sigma ^{\pm}_{j(\alpha)}= \frac {1}{2}(\sigma ^x_{j(\alpha)}
\pm i\sigma ^y_{j(\alpha)}) $ with $\sigma ^x_{j(\alpha)},
\sigma ^y_{j(\alpha)},\sigma ^z_{j(\alpha)}$ being the usual Pauli 
spin operators at site $j$  corresponding
to the $\a$-th ($\a=1,2,3$) $XY$ spin chain,
$\t(\a'-\a)$ is a step
function of $(\a'-\a)$ and $\eta$ is a coupling constant;
$B_1$ and $B_L$ are left and right boundary terms of the form
\bea
B_{1}^{(m)} &=&
\left\{\begin{array}{ll}
\frac{1}{2c_- \exp(2\eta)}\left[\begin{array}{l}\cosh ^2 \eta 
	  \sum _{\a=1}^{3}\sigma ^z_{1(\a)} \\
+\sinh \eta \cosh \eta \sum ^{3}_{\mbox{\scriptsize $\begin{array}{l}\a,\b=1\\ \a \neq 
\b\end{array}$}}\sigma ^z_{1(\a)} \sigma ^z_{1(\b)}\\
+\sinh^2 \eta\; \sigma ^z_{1(1)}
	  \sigma ^z_{1(2)} \sigma ^z _{1(3)}\end{array}\right], & \rm{for}~~m=1\\
\frac {exp(\eta)}{2c_-}\left[\cosh  \eta 
	  \sum _{\a=2}^{3}\sigma ^z_{1(\a)} 
          +\sinh \eta \sigma ^z_{1(2)} \sigma ^z_{1(3)}\right], & 
\rm{for}~~m=2\\
\frac{1}{2c_-}\sigma_{1(3)}^z, &\rm{for}~~m=3
\end{array}\right.\label{Ham-bt-1}\\
B_{L}^{(l)} &=&
\left\{\begin{array}{ll}
\frac{1}{2c_+}\left[\begin{array}{l}\cosh ^2 \eta 
	  \sum _{\a=1}^{3}\sigma ^z_{L(\a)} \\
          +\sinh \eta \cosh \eta \sum ^{3}_{\mbox{\scriptsize $\begin{array}{l}\a,\b=1\\ \a \neq 
\b\end{array}$}}
\sigma ^z_{L(\a)} \sigma ^z_{L(\b)}\\
+\sinh^2 \eta\; \sigma ^z_{L(1)}
	  \sigma ^z_{L(2)} \sigma ^z _{L(3)}\end{array}\right], & \rm{for}~~l=1\\
\frac{1}{2c_+exp(\eta)}\left[\cosh  \eta 
	  \sum _{\a=1}^{2}\sigma ^z_{L(\a)} 
          +\sinh \eta \sigma ^z_{L(1)} \sigma ^z_{L(2)}\right], & 
\rm{for}~~l=2\\
\frac{1}{2c_+}\sigma_{L(1)}^z, &\rm{for}~~l=3
\end{array}\right.\label{Ham-bt-2}
\eea
where $c_{\pm}$ are parameters describing boundary effects. It is worth mentioning 
that the boundary terms containing three- and two-spin  interactions and chemical 
potentials  at the left and right ends are consistent with the bulk symmetry 
which is a combination of the inversion $j\rightarrow L-j+1$ and the exchange among 
the sublattices. With the different choices of the pair $(m,l)~~m,l=1,2,3$, it appears 
that there exist nine classes of integrable boundary terms compatible with the 
integrability of the model.  After a generalized Jordan-Wigner transformation, 
Hamiltonian (\ref{hamiltonian}) becomes
a strongly correlated electronic system with boundary interactions.
Quite remarkably, if one restricts the Hilbert space to the
one which only consists of, say, $\s _{(1)},~\s _{(2)}$, then 
Hamiltonian ({\ref{hamiltonian}) reduces to that of two coupled
$XY$ open chains with special boundary interactions, which has been
considered in \cite{bariev}.

We now establish the quantum integrability for the system defined by
Hamiltonian (\ref{hamiltonian}), by using the general formalism 
described in the paper \cite{op4}. To fix notation,  let us 
briefly recall some basic quantities  for the bulk model (\ref{bulk-h})
with  periodic boundary conditions. As was shown in \cite{Zhou97},
the bulk model Hamiltonian  commutes with  a one-parameter 
family of bulk transfer matrix $\tau(u)$
of a two-dimensional lattice statistical mechanics model.
This transfer matrix is the trace of a monodromy
matrix $T(u)$, which is defined , as usual, by 
\begin{equation}
T(u)=L_{0N}(u)\cdots L_{01}(u) \label{TM}
\end{equation}
with $L_{0j}(u)$ of the form,
\beq
L_{0j}(u) = L^{(1)}_{0j}(u) L^{(2)}_{0j}(u) L^{(3)}_{0j}(u),\label{mono}
\eeq
where
\bea
L^{(\alpha)}_{0j}(u) &=& \frac{1}{2}(1+\sigma ^z_{j(\alpha)}
\sigma ^z_{0(\alpha)}) + \frac{1}{2} u (1-\sigma ^z_{j(\alpha)}
\sigma ^z_{0(\alpha)})
\exp(\eta \sum^{3}_{\stackrel {\alpha'=1}{\alpha' \neq \alpha}}
\sigma ^+_{0(\alpha')}
\sigma ^-_{0(\alpha')})\no\\
& &+(\sigma^-_{j(\alpha)} \sigma^+_{0(\alpha)} + \sigma ^+_{j(\alpha)}
\sigma ^-_{0(\alpha)}) \sqrt {1+
\exp(2 \eta \sum^{3}_{\stackrel {\alpha'=1}{\alpha' \neq \alpha}}
\sigma ^+_{0(\alpha')}\sigma ^-_{0(\alpha')}) u ^2}. 
\eea
The commutativity of the bulk transfer matrices
$\tau(u)$ for different values of the spectral parameter $ u$ follows
from the fact that
the monodromy matrix $T(u)$  satisfies the Yang-Baxter 
algebra
\begin{equation}
R_{12}(u,v)\stackrel{1}{T}(u)\stackrel{2}{T}(v)=
\stackrel{2}{T}(v)\stackrel{1}{T}(u)R_{12}(u,v).
\label{YBA}
\end{equation}
The explicit form of the corresponding R-matrix 
$R_{12}(u_1,u_2)$  can be found in \cite{Zhou97}.
Here we only emphasize that the local monodromy matrix as well as 
the quantum R-matrix does not possess the crossing symmetry. 
It satisfies the non-additive Yang-Baxter equation
\begin{equation}
R_{12}(u,v) R_{13}(u,w) R_{23}(v,w) =
R_{23}(v,w) R_{13}(u,w) R_{12}(u,v).
\end{equation}

Following the QISM adapted to the case of special boundary conditions,
we define the doubled monodromy matrix as
\begin{equation}
U_{-}(u)= T(u)K_{-}(u)T^{-1}(-u),
\label{DMM}
\end{equation}
such that the boundary transfer matrix is given by
\begin{equation}
\tau (u)=Tr_{0}K_{+} (u)U_{-}(u),
\label{DTM}
\end{equation}
where $T^{-1}$ is the inverse of the monodromy $T$ and
$K_{\pm}$ are the matrices defining the boundaries. The requirement
that they obey the RE \cite{bariev,zhou3}
\begin{eqnarray}
& &R_{12}(u,v)\stackrel{1}{K_-}(u)R_{21}(v,-u)\stackrel{2}{K_-}(v)\nonumber\\
& &= \stackrel{2}{K_-}(v)R_{12}(u,-v)\stackrel{1}{K_-}(u)R_{21}(-v,-u),  \label{RE1}\\
& &R_{21}^{{\rm t}_1{\rm t}_2}(v,u)\stackrel{1}{K_+^{
{\rm t}_1}}(u)\tilde{R}_{12}(-u,v)
\stackrel{2}{K_+^{{\rm t}_2}}(v)=  \nonumber \\
& &\stackrel{2}{K_+^{{\rm t}_2}}(v)\tilde{R}_{21}(-v,u)
\stackrel{1}{K_+^{{\rm t}_1}}(u) R
_{12}^{{\rm t}_1{\rm t}_2}(-u,-v)  \label{RE2}
\end{eqnarray}
together with the Yang-Baxter algebra and the following properties
\begin{eqnarray}
R_{12}(u,v)R_{21}(v,u) & = & 1,\\
\tilde{R}_{21}^{{\rm t}_1}(-v,u)R_{12}^{{\rm t}_2}(u,-v) & = & 1,\label{tildeR1}\\
\tilde{R}_{12}^{{\rm t}_2}(-u,v)R_{21}^{{\rm t}_1}(v,-u) & = & 1.\label{tildeR2}
\end{eqnarray}
assure that the transfer matrix
commutes for different spectral parameters, proving the integrability
of the model.

Therefore the transfer matrix (\ref{DTM}) may be considered as the
generating fuction of infinitely many  integrals of motion for the system.
We emphasize that there is no isomorphism between
the matrices $K_{+}(u)$ and $K_{-}(u)$,
due to the absence of the crossing symmetry
for the $R$-matrix.
Therefore, we have to solve
two REs separately in order to fix the boundaries.
In the Appendix, this calculation is presented in detail
using the {\it variable-separation} prescription.
We find three  different classes of boundary $K_\pm $ matrices
consistent with the integrability of the Hamiltonian (\ref{hamiltonian}).
Let us first list those related with  $K_-$-matrix 
\beq
K_-^{(m)}(u) = \frac {1}{\lambda _-}
\left ( \begin {array} {cccccccc}
A_-(u)  &0&0&0&0&0&0&0\\
0&B_-(u) &0&0&0&0&0&0 \\
0&0&C_-(u) &0&0&0&0&0\\
0&0&0&D_-(u)&0&0&0&0\\
0&0&0&0&E_-(u)&0&0&0\\
0&0&0&0&0&F_-(u)&0&0\\
0&0&0&0&0&0&G_-(u)&0\\
0&0&0&0&0&0&0&H_-(u)
\end {array}  \right ),\label{Km}
\eeq
where for $m=1$
\bea
A_-(u)&=&(c_-+u)(e^{2\eta} c_- +u)(e^{4\eta} c_-+u),\no\\
B_-(u)&=&(c_--u)(e^{2\eta} c_- +u)(e^{4\eta}c_-+u),\no\\
C_-(u)&=&(c_--u)(e^{2\eta} c_- +u)(e^{4\eta}c_-+u),\no\\
D_-(u)&=&(c_--u)(e^{2\eta} c_--u)(e^{4\eta} c_-+u),\no\\
E_-(u)&=&(c_--u)(e^{2\eta} c_- +u)(e^{4\eta}c_-+u),\no\\
F_-(u)&=&(c_--u)(e^{2\eta} c_--u)(e^{4\eta} c_-+u),\no\\
G_-(u)&=&(c_--u)(e^{2\eta} c_--u)(e^{4\eta} c_-+u),\no\\
H_-(u)&=&(c_--u)(e^{2\eta} c_- -u)(e^{4\eta} c_--u).\no\\
\lambda _-& =& \frac{1}{e^{6\eta}c_-^3}\no
\eea
for $m=2$
\bea
A_-(u)&=&E_-(u)=(c_-+u)(c_-+e^{2\eta}u),\no\\
B_-(u)&=&C_-(u)=F_-(u)=G_-(u)=(c_-+u)(c_--e^{2\eta}u),\no\\
D_-(u)&=&H_-(u)=(c_--u)(c_--e^{2\eta}u),\no\\
\lambda _-& =& \frac{1}{c_-^2}\no
\eea
for $m=3$
\bea
A_-(u)&=&C_-(u)=E_-(u)=G_-(u)=(c_-+u),\no\\
B_-(u)&=&D_-(u)=F_-(u)=H_-(u)=(c_--u),\no\\
\lambda _-& =& \frac{1}{c_-}\no
\eea

However, it is much more tedious to find the boundary K-matrix
$K_+(u)$, since not only the corresponding RE is more involved but
also the new object $\tilde{R}_{12}(u,v)$ is more complicated. We list
the final result here and some details can be found in Appendix, \beq
K_+^{(l)}(u)= \left ( \begin {array} {cccccccc} A_+(u)
&0&0&0&0&0&0&0\\ 0&B_+(u) &0&0&0&0&0&0 \\ 0&0&C_+(u) &0&0&0&0&0\\
0&0&0&D_+(u)&0&0&0&0\\ 0&0&0&0&E_+(u)&0&0&0\\ 0&0&0&0&0&F_+(u)&0&0\\
0&0&0&0&0&0&G_+(u)&0\\ 0&0&0&0&0&0&0&H_+(u)
\end {array}  \right ) ,   \label{Kp}
\eeq for $l=1$ \bea A_+(u)&=&(e^{6\eta}c_+u-1)(e^{4\eta}c_+u -1)
(e^{2\eta}c_+u-1) ,\no\\
B_+(u)&=&e^{4\eta}(e^{2\eta}c_+u+1)(e^{4\eta}c_+u -1)
(e^{2\eta}c_+u-1) ,\no\\
C_+(u)&=&e^{2\eta}(e^{2\eta}c_+u+1)(e^{4\eta}c_+u -1)
(e^{2\eta}c_+u-1),\no\\ D_+(u)&=&e^{6\eta}(e^{2\eta}c_+u+1)(c_+u +1)
(e^{2\eta}c_+u-1),\no\\ E_+(u)&=&(e^{2\eta}c_+u+1)(e^{4\eta}c_+u -1)
(e^{2\eta}c_+u-1) ,\no\\ F_+(u)&=&e^{4\eta}(c_+u+1)(e^{2\eta}c_+u +1)
(e^{2\eta}c_+u-1) ,\no\\ G_+(u)&=&e^{2\eta}(c_+u+1)(e^{2\eta}c_+u +1)
(e^{2\eta}c_+u-1),\no\\ H_+(u)&=&e^{4\eta}(c_+u+e^{2\eta})(c_+u +1)
(e^{2\eta}c_+u+1).\no \eea for $l=2$ \bea
A_+(u)&=&B_+(u)=(e^{6\eta}c_+u-1)(e^{4\eta}c_+u -1),\no\\
C_+(u)&=&D_+(u)=e^{2\eta}(e^{2\eta}c_+u+1)(e^{4\eta}c_+u -1),\no\\
E_+(u)&=&F_+(u)=(e^{2\eta}c_+u+1)(e^{4\eta}c_+u -1),\no\\
G_+(u)&=&H_+(u)=e^{2\eta}(c_+u+1)(e^{2\eta}c_+u +1),\no \eea for $l=3$
\bea A_+(u)&=&B_+(u)=e^{2\eta}(e^{4\eta}c_+u -1),\no\\
C_+(u)&=&D_+(u)=(e^{4\eta}c_+u -1),\no\\
E_+(u)&=&F_+(u)=e^{2\eta}(c_+u +1),\no\\ G_+(u)&=&H_+(u)=(c_+u+1),\no
\eea The above explicit formulae for $K_\pm(u)$, derived by solving
the two REs directly, clearly show that no automorphism between
$K_+(u)$ and $K_-(u)$ exists and $K_+(u)$ can not be obtained from
$K_-(u)$.  These three classes of boundary
$K_{\pm }$-matrices provide nine possible choices of BC, according to
the combination of the boundary pairs
($K_{-}^{(m)}(u),K_{+}^{(l)}(u)$), $m,l=1,2,3$, which originate the pair-boundary
terms ($B_1^{(m)}$, $B_L^{(l)}$), respectively. Taking into account
the fact that $TrK_{+}^{(1)}(0)=0$ , $Tr\stackrel{\bullet}{K}_{+}^{(1)}(0)=0$,
$Tr\stackrel{\bullet \bullet}{K}_{+}^{(1)}(0)=0$ and $TrK_{+}^{(2)}(0)=0$,
$Tr\stackrel{\bullet}{K}_{+}^{(2)}(0)=0$, as well as $TrK_{+}^{(3)}(0)=0$,
 we can
show that the Hamiltonian (\ref{hamiltonian}) with the boundary terms
pairs ($K_{-}^{(m)}(u),K_{+}^{(l)}(u)$) is related to the
transfer matrix matrix (\ref{DTM}) in the following way 
\bea \tau(u) &
= & C_{1} u^3+C_{2}(H+{\rm const.})~u^4+\cdots,~~~~{\rm for}~~~l=1,
\label{H-DTM-1}\\ \tau (u) & = & C_{3}u^2+C_{4}(H+{\rm
const.})~u^{3}+\cdots,~~~~{\rm for}~~~l=2,\label{H-DTM-2}\\ \tau(u) &
= & C_{5}u+C_{6}(H+{\rm const.})~u^{2}+\cdots,~~~~{\rm
for}~~~l=3.\label{H-DTM-3} \eea Above $C_{i},\,i=1,\cdots 6,$ are some
scalar functions of the boundary parameters. The symbols, either ``bullet'' or ``prime'' denote 
the
derivative with respect to the spectral parameter.  For $l=1$, the Hamiltonian
(\ref{hamiltonian}) is related to the fourth derivative of the
boundary transfer matrix $\tau(u)$ which  can be derived by the  simplified
formula \cite{zhou3} 
\bea H & \equiv & \frac{\tau ^{''''}(0)}{8 {\rm
Tr}{K_+^{(1)}}^{'''}(0)}= \sum ^{L-1}_{j=1} H_{jj+1}
+\frac{1}{2}P_{01}{K^{(m)}_-}^{'}(0)P_{01}\no \\ & & +\frac{3}{{\rm
Tr}{K_+^{(1)}}^{'''}(0)}\;{\rm
Tr}({K_+^{(1)}}^{''}(0)L_{0L}^{'}(0)P_{0L}),\label{4-th-der} \\
H_{j,j+1} & = & P_{0,j+1}(0)L'_{0j}(0)P^{-1}_{0j}(0)P^{-1}_{0,j+1}(0)
\eea with $m=1,2,3$. 

However, to obtain new boundary terms
corresponding to the $K_+^{(l)}$-matrices when $l=2,3$, we have to
develop new formulae regarding to the relations (\ref{H-DTM-2}) and
(\ref{H-DTM-3}).  What follows are the general formulae of the
Hamiltonian related to Eqs (\ref{H-DTM-2}) and (\ref{H-DTM-3}) given
by \bea H & \equiv & \frac{t^{'''}(0)}{f}= \sum ^{L-1}_{j=1} H_{jj+1}
+\frac{1}{2}P_{01}{K^{(m)}_-}^{'}(0)P_{01}\no \\ & &
+\frac{1}{f}\left\{ {\rm Tr}({K_+^{(2)}}(0)L_{0L}^{'''}(0)P_{0L})
+{\rm Tr}({K_+^{(2)}}(0)P_{0L}\bar{L}_{0L}^{'''}(0))\right.\no\\ & &
\left.+6\;{\rm Tr}({K_+^{(2)}}^{''}(0)L_{0L}^{'}(0)P_{0L}) +3\;{\rm
Tr}({K_+^{(2)}}^{'}(0)L_{0L}^{''}(0)P_{0L})\right.\no\\ & &
\left.+3\;{\rm Tr}({K_+^{(2)}}^{'}(0)P_{0L}\bar{L}_{0L}^{''}(0))
+3\;{\rm
Tr}({K_+^{(2)}}(0)L_{0L}^{'}(0)\bar{L}_{0L}^{''}(0))\right.\no\\ & &
\left.+3\;{\rm Tr}({K_+^{(2)}}(0)L_{0L}^{''}(0)\bar{L}^{'}_{0L}(0))
+6\;{\rm
Tr}({K_+^{(2)}}^{'}(0)L_{0L}^{'}(0)\bar{L}^{'}_{0L}(0))\right\}, \eea
and \bea H & \equiv & \frac{t^{(2)}(0)}{{\rm Tr}{K_+^{(3)}}^{'}(0)}
=\sum ^{L-1}_{j=1} H_{jj+1}
+\frac{1}{2}P_{01}{K^{(m)}_-}^{'}(0)P_{01}\no \\ & & + \frac{1}{{\rm
Tr}{K_+^{(3)}}^{'}(0)}\left\{ {\rm
Tr}({K_+^{(2)}}(0)L_{0L}^{''}(0)P_{0L})+{\rm
Tr}({K_+^{(2)}}(0)P_{0L}\bar{L}_{0L}^{''}(0))\right.\no \\ & &
\left.+2\;{\rm Tr}({K_+^{(2)}}(0)L_{0L}^{'}(0)\bar{L}^{'}_{0L}(0))
+2\;{\rm Tr}({K_+^{(2)}}^{'}(0)L_{0L}^{'}(0)P_{0L})\right\}, \eea
respectively.  Above \bea & &\bar{L}_{0L}^{'}(0)\equiv
\frac{d}{d\;u}L^{-1}_{0L}(-u)|_{u=0}=P_{0L}L^{'}(0)P_{0L},\no\\ &
&\bar{L}_{0L}^{''}(0) \equiv \frac{d^2}{d\;u^2}L^{-1}_{0L}(-u)|_{u=0}
\neq P_{0L}L^{''}(0)P_{0L},\no\\ & &\bar{L}_{0L}^{'''}(0)\equiv
\frac{d^3}{d\;u^3}L^{-1}_{0L}(-u)|_{u=0} \neq
P_{0L}L^{''}(0)P_{0L},\no\\ & &f = 6\;\left[{\rm
Tr}({K_+^{(2)}}(0)L_{0L}^{''}(0)P_{0L}) +{\rm
Tr}({K_+^{(2)}}(0)P_{L0}\bar{L}_{0L}^{''}(0))+ {\rm
Tr}{K_+^{(2)}}^{''}(0)\right].  \eea 
Using above formulae and after
lengthy and tough calculation, one can derive the boundary terms
presented by Eqs (\ref{Ham-bt-1}) and (\ref{Ham-bt-2}).  It was seen
that the Hamiltonian of the model can be derived from expansion of the
boundary transfer matrix (\ref{DTM}) at $u=0$ up to the fourth, third
and second orders respectively. This contrasts to the case of two
coupled XY spin chain where the Hamiltonian with four classes of the
boundary terms coming from the third and second derivatives of the
transfer matrix \cite{bariev}. In the next
sections, we shall be focusing on the solution of the eigenvalue
problem of the Hamiltonian (\ref{hamiltonian}).

\sect{The Bethe ansatz equations \label{bethe}}

Having established the quantum integrability of the model,
let us  now  solve it by using
the coordinate space Bethe ansatz method. Following
\cite{Asak}, we
assume that the eigenfunction of the Hamiltonian (\ref{hamiltonian}) 
takes the form
\bea
| \Psi \rangle &  = & \sum _{\{(x_j,\a_j)\}}\Psi _{\a_1,\cdots,\a_N}\s^+
_{x_1\a	_1}\cdots \s ^+_{x_N\a_N} | 0 \rangle,\\
\Psi_{\a_1,\cdots,\a_N}(x_1,\cdots,x_N)
&=&\sum _P \e _P A_{\a_{Q1},\cdots,\a_{QN}}(k_{PQ1},\cdots,k_{PQN})
\exp (i\sum ^N_{j=1} k_{P_j}x_j),\no
\eea
where the summation is taken over all permutations and negations of
$k_1,\cdots,k_N,$ and $Q$ is the permutation of the $N$ particles such that
$1\leq   x_{Q1}\leq   \cdots  \leq  x_{QN}\leq   L$.
The symbol $\e_P$ is a sign factor $\pm1$ and changes its sign
under each 'mutation'. Substituting the wavefunction into  the
eigenvalue equation $ H| \Psi  \rangle = E | \Psi \rangle $,
one gets
\bea
A_{\cdots,\a_j,\a_i,\cdots}(\cdots,k_j,k_i,\cdots)&=&S_{ij}(
k_i,k_j)
    A_{\cdots,\a_i,\a_j,\cdots}(\cdots,k_i,k_j,\cdots),\no\\
A_{\a_i,\cdots}(-k_j,\cdots)&=&s^L(k_j;p_{1\a_i})A_{\a_i,\cdots}
    (k_j,\cdots),\no\\
A_{\cdots,\a_i}(\cdots,-k_j)&=&s^R(k_j;p_{L\a_i})A_{\cdots,\a_i}(\cdots,k_j)
\eea
with $S_{ij}(k_i,k_j) \equiv S_{ij}(\frac {1}{2}(k_i-k_j))$ being
the two-particle scattering matrix,
\bea
S^{\a \a}_{\a \a}(k)& = &1,~~~ S^{\a \b}_{\a \b}(k)
 =\frac{\sin k}{\sin (k-i \eta)},\no\\
S^{\b \a}_{\a \b}(k)& = &-i e^{i\;{\rm sgn} (\b -\a)k}
  \frac{\sinh\eta}{\sin (k-i\eta)},~~~~
  \a \neq \b ,
\eea
and 
$s^L(k_j;p_{1\a_i})$
and $s^R(k_j;p_{L\a_i})$ the boundary scattering matrices,
\bea
s^L(k_j;p_{1\a_i})&=&\frac {1-p_{1\a_i}e^{ik_j}}
{1-p_{1\a_i}e^{-ik_j}},\no\\
s^R(k_j;p_{L\a_i})&=&\frac {1-p_{L\a_i}e^{-ik_j}}
{1-p_{L\a_i}e^{ik_j}}e^{2ik_j(L+1)}\label{sm}
\eea
where $\a_i=1,2,3$ and  
$p_{1\a_i}$ and $p_{L\a_i}$ are given by the following formulae,
corresponding to the nine cases with respect to the boundary pair ($B_1^{(m)},B_L^{(l)}$), 
respectively,
\bea
{\rm Case~i:}~~~
& &p_{11}=p_{12}=p_{13}\equiv p_1=
   \frac{1}{c_-\exp(4\eta)},\no\\
&&p_{L1}=p_{L2}=p_{L3}\equiv p_L=
  \frac{1}{c_+\exp(2\eta)};\label {p1}\\
{\rm Case~ii:}~~
& &p_{11}=0,~~~p_{12}=p_{13}\equiv p_{1+}=
   \frac{1}{c_-},\no\\
&&p_{L1}=p_{L2}\equiv p_{L+}= \frac{1}{c_+ \exp(2\eta)},~~~
p_{L3}=0 ;\label {p2}\\
{\rm Case~iii:}~~
& &p_{11}=p_{12}=0,~~~p_{13}= \frac{1}{c_-},\no\\
&&p_{L1}=\frac{1}{c_+}, ~~~p_{L2}=p_{L3}=
  0;\label {p3}\\
{\rm Case~iv:}~~
& &p_{11}=p_{12}=p_{13}\equiv p_1=
   \frac{1}{c_-\exp(4\eta)},\no\\
&&p_{L1}=p_{L2}= \frac{1}{c_+ \exp(2\eta)},~~~
p_{L3}=0 ;\label {p4}\\
{\rm Case~v:}~~
& &p_{11}=p_{12}=p_{13}\equiv p_1=
   \frac{1}{c_-\exp(4\eta)},\no\\
&&p_{L1}=\frac{1}{c_+}, ~~~p_{L2}=p_{L3}=
  0;\label {p5}\\
{\rm Case~vi:}~~
& &p_{11}=0,~~~p_{12}=p_{13}=
   \frac{1}{c_-},\no\\
&&p_{L1}=\frac{1}{c_+}, ~~~p_{L2}=p_{L3}=
  0;\label {p6}\\
{\rm Case~vii:}~~
& &p_{11}=0,~~~p_{12}=p_{13}=
   \frac{1}{c_-},\no\\
&&p_{L1}=p_{L2}=p_{L3}\equiv p_L=
  \frac{1}{c_+\exp(2\eta)};\label {p7}\\
{\rm Case~viii:}~~
& &p_{11}=p_{12}=0,~~~p_{13}= \frac{1}{c_-},\no\\
&&p_{L1}=p_{L2}=p_{L3}\equiv p_L=
  \frac{1}{c_+\exp(2\eta)};\label {p8}\\
{\rm Case~ix:}~~
& &p_{11}=p_{12}=0,~~~p_{13}= \frac{1}{c_-},\no\\
&&p_{L1}=p_{L2}= \frac{1}{c_+ \exp(2\eta)},~~~
p_{L3}=0 ;\label {p9}
\eea

As is seen above, the two-particle S-matrix (\ref{sm}) is 
nothing but the $R$ matrix of the $A_2$ XXZ Heisenberg model (in the homogeneous
gauge) and thus satisfies the quantum Yang-Baxter equation,
and the boundary scattering matrices $s^L$ and $s^R$ 
obey the corresponding reflection equations (for details, see, \cite{deV93}).
This is seen as follows. One introduces the notation
\beq
s(k;p)=\frac  {1-pe^{-ik}}{1-p e^{ik}}.
\eeq
Then the boundary scattering matrices $s^L(k_j;p_{1\a_i})$,
 $~s^R(k_j;p_{L\a_i})$ can be written as, corresponding to the nine
cases, respectively,
\bea
& &{\rm Case~i:}~~s^L(k_j;p_{1\a_i})=s(-k_j;p_1)I,\no\\
& &s^R(k_j;p_{L\a_i})=e^{ik_j2(L+1)}s(k_j;p_L)I;\label{s11}\\
& &{\rm Case~ii:}~~s^L(k_j;p_{1\a_i})=s(-k_j;p_{1+})\lt(
\begin{array}{ccc}
1 &0 &0\\
0&e^{ik_j} \frac{\sin (i \zeta_-+k_j/2)}
	       {\sin (i \zeta_--k_j/2)} & 0\\
0&0&e^{ik_j} \frac{\sin (i \zeta_-+k_j/2)}
	       {\sin (i \zeta_--k_j/2)}
\end{array}
\rt),\no\\
& &s^R(k_j;p_{L\a_i})=e^{ik_j2(L+1)}s(k_j;p_{L1})\lt(
\begin{array}{ccc}
1&0&0\\
0&1&0\\
0&0&e^{ik_j} \frac{\sin (i\k_++k_j/2)}
	       {\sin (i\k_+-k_j/2)}
\end{array}\rt);\label{s22}\\
& &{\rm Case~iii:}~~s^L(k_j;p_{1\a_i})=
s(-k_j;p_{11})\lt(
\begin{array}{ccc}
1 &0 &0\\
0&1 & 0\\
0&0&e^{ik_j} \frac{\sin (i\k_-+k_j/2)}
	       {\sin (i\k_--k_j/2)}
\end{array}
\rt),\no\\
& &s^R(k_j;p_{L\a_i})=e^{ik_j2(L+1)}s(k_j;p_{L1})\lt(
\begin{array}{ccc}
1&0&0\\
0&e^{ik_j}\frac{\sin (i \zeta_++k_j/2)}
	       {\sin (i \zeta _+-k_j/2)}&0\\ 
0&0&e^{ik_j}\frac{\sin (i \zeta_++k_j/2)}
	       {\sin (i \zeta _+-k_j/2)} 
\end{array}\rt);\label{s33}\\
& &{\rm Case~iv:}~~s^L(k_j;p_{1\a_i})=
s(-k_j;p_1)I,\no\\
& &s^R(k_j;p_{L\a_i})=e^{ik_j2(L+1)}s(k_j;p_{L1})\lt(
\begin{array}{ccc}
1&0&0\\
0&1&0\\
0&0&e^{ik_j} \frac{\sin (i\k_++k_j/2)}
	       {\sin (i\k_+-k_j/2)}
\end{array}
\rt);\label{s12}\\
& &{\rm Case~v:}~~s^L(k_j;p_{1\a_i})=
s(-k_j;p_1)I,\no\\
& &s^R(k_j;p_{L\a_i})=e^{ik_j2(L+1)}s(k_j;p_{L1})\lt(
\begin{array}{ccc}
1&0&0\\
0&e^{ik_j}\frac{\sin (i \zeta_++k_j/2)}
	       {\sin (i \zeta _+-k_j/2)}&0\\ 
0&0&e^{ik_j}\frac{\sin (i \zeta_++k_j/2)}
	       {\sin (i \zeta _+-k_j/2)} 
\end{array}\rt);\label{s13}\\
& &{\rm Case~vi:}~~s^L(k_j;p_{1\a_i})=s(-k_j;p_{1+})\lt(
\begin{array}{ccc}
1 &0 &0\\
0&e^{ik_j} \frac{\sin (i \zeta_-+k_j/2)}
	       {\sin (i \zeta_--k_j/2)} & 0\\
0&0&e^{ik_j} \frac{\sin (i \zeta_-+k_j/2)}
	       {\sin (i \zeta_--k_j/2)}
\end{array}
\rt),\no\\
& &s^R(k_j;p_{L\a_i})=e^{ik_j2(L+1)}s(k_j;p_{L1})\lt(
\begin{array}{ccc}
1&0&0\\
0&e^{ik_j}\frac{\sin (i \zeta_++k_j/2)}
	       {\sin (i \zeta _+-k_j/2)}&0\\ 
0&0&e^{ik_j}\frac{\sin (i \zeta_++k_j/2)}
	       {\sin (i \zeta _+-k_j/2)} 
\end{array}
\rt);\label{s23}\\
& &{\rm Case~vii:}~~s^L(k_j;p_{1\a_i})=s(-k_j;p_{1+})\lt(
\begin{array}{ccc}
1 &0 &0\\
0&e^{ik_j} \frac{\sin (i \zeta_-+k_j/2)}
	       {\sin (i \zeta_--k_j/2)} & 0\\
0&0&e^{ik_j} \frac{\sin (i \zeta_-+k_j/2)}
	       {\sin (i \zeta_--k_j/2)}
\end{array}
\rt),\no\\
& &s^R(k_j;p_{L\a_i})=e^{ik_j2(L+1)}s(k_j;p_L)I;\label{s21}\\
& &{\rm Case~viii:}~~s^L(k_j;p_{1\a_i})=
s(-k_j;p_{11})\lt(
\begin{array}{ccc}
1 &0 &0\\
0&1 & 0\\
0&0&e^{ik_j} \frac{\sin (i\k_-+k_j/2)}
	       {\sin (i\k_--k_j/2)}
\end{array}
\rt),\no\\
& &s^R(k_j;p_{L\a_i})=e^{ik_j2(L+1)}s(k_j;p_L)I;\label{s31}\\
& &{\rm Case~ix:}~~s^L(k_j;p_{1\a_i})=
s(-k_j;p_{11})\lt(
\begin{array}{ccc}
1 &0 &0\\
0&1 & 0\\
0&0&e^{ik_j} \frac{\sin (i\k_-+k_j/2)}
	       {\sin (i\k_--k_j/2)}
\end{array}
\rt),\no\\
& &s^R(k_j;p_{L\a_i})=e^{ik_j2(L+1)}s(k_j;p_{L1})\lt(
\begin{array}{ccc}
1&0&0\\
0&1&0\\
0&0&e^{ik_j} \frac{\sin (i\k_++k_j/2)}
	       {\sin (i\k_+-k_j/2)}
\end{array}
\rt);\label{s32}
\eea
Here $I$ stands for $3\times 3$ identity
matrix and $p_{1+},~p_{L+}$ are the ones given in (\ref{p2}); $\zeta
_{\pm}, \k _{\pm}$ are parameters defined by \beq e^{2\zeta _{\pm}} =
c_{\pm}, ~~~ e^{2\k_+} = c_+ e^{2\eta}, ~~~ e^{2\k_-} = c_-.  \eeq We
immediately see that (\ref{s11}) are the trivial solutions of the
reflection equations, whereas (\ref{s22}) and (\ref{s33}) are the
diagonal solutions for the $A_2$ $XXZ$ model $R$ matrix. The boundary
scattering matrices given in (\ref{s11}), (\ref{s22}) and (\ref{s33})
also constitute nine classes of possible choices of boundary
conditions for the spin degrees of freedom of the model.

Then, the diagonalization of Hamiltonian (\ref{hamiltonian}) reduces 
to solving  the following matrix  eigenvalue equation
\beq
T_jt= t,~~~~~~~j=1,\cdots,N,
\eeq
where $t$ denotes an eigenvector on the space of the spin variables
and $T_j$ takes the form
\beq
T_j=S_j^-(k_j)s^L(-k_j;p_{1\a_j})R^-_j(k_j)R^+_j(k_j)
    s^R(k_j;p_{L\a_j})S^+_j(k_j)
\eeq
with
\bea
S_j^+(k_j)&=&S_{j,N}(k_j,k_N) \cdots S_{j,j+1}(k_j,k_{j+1}),\no\\
S^-_j(k_j)&=&S_{j,j-1}(k_j,k_{j-1})\cdots S_{j,1}(k_j,k_1),\no\\
R^-_j(k_j)&=&S_{1,j}(k_1,-k_j)\cdots S_{j-1,j}(k_{j-1},-k_j),\no\\
R^+_j(k_j)&=&S_{j+1,j}(k_{j+1},-k_j)\cdots S_{N,j}(k_N,-k_j).
\eea
This problem may be solved using the algebraic Bethe ansatz method.
The Bethe ansatz equations are 
\bea
 & &e^{ik_j2(L+1)} F(k_j; p_{11},p_{L1})  
= \prod ^{M_1}_{\a =1}\frac {\sin [\frac {1}{2}
(k_j-\Lambda ^{(1)}_{\a})+\frac {i\eta}{2}]}
{\sin [\frac {1}{2}
(k_j-\Lambda ^{(1)}_{\a}) -\frac {i\eta}{2}]}
\frac {\sin [\frac {1}{2}
(k_j+\Lambda ^{(1)}_{\a})+\frac {i\eta}{2}]}
{\sin [\frac {1}{2}
(k_j+\Lambda ^{(1)}_{\a}) -\frac {i\eta}{2}]},\no\\
&  &\prod ^N_{\a =1}\frac {\sin [\frac {1}{2}(\Lambda ^{(1)}_{\g}
  -k_\a)+\frac{i\eta}{2}]}
{\sin [\frac {1}{2}(\Lambda ^{(1)}_{\g}
   -k_\a)-\frac{i\eta}{2}]}
\frac {\sin [\frac {1}{2}(\Lambda ^{(1)}_{\g}
   +k_\a+\frac{i\eta}{2}]}
{\sin [\frac {1}{2}(\Lambda ^{(1)}_{\g}
   +k_\a)-\frac{i\eta}{2}]}\no \\
& &
=G(\Lambda ^{(1)}_{\g}; \zeta_-, \zeta_+) 
 \prod ^{M_1}_{\stackrel {\g ' =1}{\g ' \neq \g}}\frac
 {\sin [\frac {1}{2}(\Lambda ^{(1)}_{\g}-\Lambda ^{(1)}_{\g '})
 +i\eta]}{\sin [\frac {1}{2}(\Lambda ^{(1)}_{\g}-\Lambda ^{(1)}_{\g '})
 -i\eta]}\frac {\sin [\frac {1}{2}(\Lambda ^{(1)}_{\g}+\Lambda ^{(1)}_{\g '})
 +i\eta]}{\sin [\frac {1}{2}(\Lambda ^{(1)}_{\g}+\Lambda ^{(1)}_{\g '})
 -i\eta]}\no\\
 & &\times\prod ^{M_2}_{\delta=1}
 \frac {\sin [\frac {1}{2}(\Lambda ^{(1)}_{\g}-\lambda ^{(2)}_{\delta})
 -\frac{i\eta}{2}]}
 {\sin [\frac {1}{2}(\Lambda ^{(1)}_{\g}-\lambda ^{(2)}_{\delta})
 +\frac{i\eta}{2}]}
 \frac {\sin [\frac {1}{2}(\Lambda ^{(1)}_{\g}+\lambda ^{(2)}_{\delta})
 -\frac{i\eta}{2}]}
 {\sin [\frac {1}{2}(\Lambda ^{(1)}_{\g}+\lambda ^{(2)}_{\delta})
 +\frac{i\eta}{2}]},\no\\
& &
\prod ^{M_2}_{\stackrel {\g '=1}{\g ' \neq \g}}
\frac {\sin [\frac {1}{2}(\Lambda ^{(2)}_{\g}-\Lambda ^{(2)}_{\g '})
 +i\eta]}{\sin [\frac {1}{2}(\Lambda ^{(2)}_{\g}-\Lambda ^{(2)}_{\g '})
 -i\eta]}\frac {\sin [\frac {1}{2}(\Lambda ^{(2)}_{\g}+\Lambda ^{(2)}_{\g '})
 +i\eta]}{\sin [\frac {1}{2}(\Lambda ^{(2)}_{\g}+\Lambda ^{(2)}_{\g '})
  -i\eta]}=
  K( \Lambda ^{(2)}_{\g}; \k_-, \k_+) \no\\
& &
\prod ^{M_1}_{\a =1} 
 \frac {\sin [\frac {1}{2}(\Lambda ^{(2)}_{\g}-\Lambda ^{(1)}_{\a})
 +\frac{i\eta}{2}]}
 {\sin [\frac {1}{2}(\Lambda ^{(2)}_{\g}-\Lambda ^{(1)}_{\a})
 -\frac{i\eta}{2}]}
 \frac {\sin [\frac {1}{2}(\Lambda ^{(2)}_{\g}+\Lambda ^{(1)}_{\a})
 +\frac{i\eta}{2}]}
 {\sin [\frac {1}{2}(\Lambda ^{(2)}_{\g}+\Lambda ^{(1)}_{\a})
 -\frac{i\eta}{2}]},
\eea
where 
\bea
F(k_j;p_{1+},p_{L+})&=& s(k_j;p_{1+}) s(k_j;p_{L+}),\,\,\,({\rm for\; all\; cases})      \no\\
G(\Lambda^{(1)}_\g;\zeta _-,\zeta _+)&=& \left \{ \begin {array}{ll}
1 & case \;(i)\\
\frac {\sin (i \zeta _- -i \Lambda^{(1)}-\g/2 + \eta/2)}
 {\sin (i \zeta _- +i \Lambda^{(1)}-\g/2 + \eta/2)} e^{i \Lambda^{(1)}_\g} 
& case\;( ii)\\
\frac {\sin (i \zeta _+ -i \Lambda^{(1)}-\g/2 + \eta/2)}
 {\sin (i \zeta _+ +i \Lambda^{(1)}-\g/2 + \eta/2)} e^{-i \Lambda^{(1)}_\g}
& case \;(iii)\\
1 & case \;(iv)\\
\frac {\sin (i \zeta _+ -i \Lambda^{(1)}-\g/2 + \eta/2)}
 {\sin (i \zeta _+ +i \Lambda^{(1)}-\g/2 + \eta/2)} e^{-i \Lambda^{(1)}_\g}
& case \;(v)\\
\frac {\sin (i \zeta _- -i \Lambda^{(1)}-\g/2 + \eta/2)}
 {\sin (i \zeta _- +i \Lambda^{(1)}-\g/2 + \eta/2)}
\frac {\sin (i \zeta _+ -i \Lambda^{(1)}-\g/2 + \eta/2)}
 {\sin (i \zeta _+ +i \Lambda^{(1)}-\g/2 + \eta/2)}
& case \;(vi)\\
\frac {\sin (i \zeta _- -i \Lambda^{(1)}-\g/2 + \eta/2)}
 {\sin (i \zeta _- +i \Lambda^{(1)}-\g/2 + \eta/2)} e^{i \Lambda^{(1)}_\g} 
 & case \;(vii)\\
1 & case \;(viii)\\
1 & case\;( ix)
\end {array} \right.\no\\
K(\Lambda^{(2)}_\g;\k_-,\k_+)&=& \left \{ \begin {array}{ll}
1 & case\;( i)\\
\frac {\sin (i \k _+ -i \Lambda^{(2)}-\g/2 + \eta)}
 {\sin (i \k _+ +i \Lambda^{(2)}-\g/2 + \eta)} e^{-i \Lambda^{(2)}_\g}
& case \;(ii)\\
\frac {\sin (i \k _- -i \Lambda^{(2)}-\g/2 + \eta)}
 {\sin (i \k _- +i \Lambda^{(2)}-\g/2 + \eta)} e^{i \Lambda^{(2)}_\g}
& case \;(iii)\\
\frac {\sin (i \k _+ -i \Lambda^{(2)}-\g/2 + \eta)}
 {\sin (i \k _+ +i \Lambda^{(2)}-\g/2 + \eta)} e^{-i \Lambda^{(2)}_\g}
& case \;(iv)\\
1 & case \;(v)\\
1 & case \;(vi)\\
1 & case \;(vii)\\
\frac {\sin (i \k _- -i \Lambda^{(2)}-\g/2 + \eta)}
 {\sin (i \k _- +i \Lambda^{(2)}-\g/2 + \eta)} e^{i \Lambda^{(2)}_\g}
& case \;(viii)\\
\frac {\sin (i \k _- -i \Lambda^{(2)}-\g/2 + \eta)}
 {\sin (i \k _- +i \Lambda^{(2)}-\g/2 + \eta)} 
\frac {\sin (i \k _+ -i \Lambda^{(2)}-\g/2 + \eta)}
 {\sin (i \k _+ +i \Lambda^{(2)}-\g/2 + \eta)} 
& case \;(ix)
\end {array} \right.\no\\
\eea
The energy eigenvalue $E$ of the model is given by
$E=-2\sum ^N_{j=1}\cos k_j$ (modular an unimportant additive constant).

\sect{Conclusion \label{con}}

In this paper, we have studied integrable open-boundary conditions for
the three coupled $XY$ spin chain model. The quantum integrability of
the boundary system has been established by the fact that the
corresponding Hamiltonian may be embedded into a one-parameter family
of commuting transfer matrices. A desirable way to handle open
boundary conditions for the models associated with a general class of
quantum R-matrices (with or  without crossing-unitarity) has been
developed.  Moreover, the Bethe Ansatz equations are derived by means
of the coordinate Bethe ansatz approach. This provides a basis for
computing the finite size corrections to the low-lying energies in the
system, which in turn could be used together with the boundary
conformal field theory technique to study the critical properties of
the boundaries.

Lastly, it is interesting to formulate a graded version of the
quantum Yang-Baxter algebra and reflection equation algebra for the fermionic
Bariev model. The algebraic Bethe ansatz for the three coupled $XY$ model
with both periodic and open boundary conditions would be very significant
in understanding of the symmetry structure of the model. 
Those will be addressed in a future publication.

\begin{ack}
A.F., I.R. and X.W.G. thank CNPq (Conselho Nacional de Desenvolvimento
Cient\'{\i}fico e Tecnol\'ogico) and FAPERGS (Funda\c{c}\~{a}o de
Amparo \~{a} Pesquisa do Estado do Rio Grande do Sul) for financial
support, I.R. also thanks PRONEX.  M.D.G. and H.Q.Z. acknowledges
the support from the Australian Research
Council. We thank R. Mckenzie and Jon Links for helpful comments on 
the manuscript.

\end{ack}

\appendix

\sect{Derivation of the boundary $K_{\pm}$-matrices}

The R-matrix is a $64\times 64$ matrix with 216 non-zero elements which are given in 
\cite{Zhou97}. We now are looking for 
  diagonal solutions $K_{\pm}(u)$ of the REs. We parametrize
$K_{\pm}(u)$ as
\beq
K_-(u) =\left (
\begin {array} {cccccccc}
z1_{\pm}(u)  &0&0&0&0&0&0&0\\
0&z2_{\pm}(u) &0&0&0&0&0&0 \\
0&0&z_{\pm}3(u) &0&0&0&0&0\\
0&0&0&z4_{\pm}(u)&0&0&0&0\\
0&0&0&0&z5_{\pm}(u)&0&0&0\\
0&0&0&0&0&z6_{\pm}(u)&0&0\\
0&0&0&0&0&0&z7_{\pm}(u)&0\\
0&0&0&0&0&0&0&z8_{\pm}(u)
\end {array}  \right ).\label{Kpm}
\eeq Using some simpler functional equations from the first  RE wich are given
by (A.1)-(A.8) in the Appendix A of the paper \cite{zhou3}, one can present
an ansatz 
\bea &&\frac{z2_-(u)}{z1_-(u)}=\frac{c_1-c_8u}{c_1+c_8u},~~~~~
\frac{z3_-(u)}{z1_-(u)}=\frac{c_2-c_9u}{c_2+c_9u},\\ & &
\frac{z5_-(u)}{z1_-(u)}=\frac{c_3-c_{10}u}{c_3+c_{10}u},~~~~~
\frac{z4_-(u)}{z2_-(u)}=\frac{c_4-c_{11}u}{c_4+c_{11}u},\\ & &
\frac{z6_-(u)}{z2_-(u)}=\frac{c_5-c_{12}u}{c_5+c_{12}u},~~~~~
\frac{z7_-(u)}{z3_-(u)}=\frac{c_6-c_{13}u}{c_6+c_{13}u},\\ & &
\frac{z8_-(u)}{z4_-(u)}=\frac{c_7-c_{14}u}{c_7+c_{14}u} 
\eea 
with minimal
coefficients $c_i$ to be determined.  Running the RE again with above
ansatz, it can be  found that only one coefficient is free and  
three classes of boundary $K_-$-matrices can be immediately chosen as
the forms presented in (\ref{Km}). 

To solve the second RE is rather cumbersome and sophisticated. For
clarifying the functional equations arising from the second RE, we
denote the Boltzmann weights associated with the $R$-matrix as
$w_1(u,v),\cdots,w_{58}(u,v)$ in accordance with the orders listed in the paper
\cite{Zhou97}. The  convenient notations  $\hat{z}=z_+(v)$ and $\bar{z}=z_+(u)$
will be implied hereafter.   Similarly,
after substituting $K_+$-matrix (\ref{Kpm}) into the RE (\ref{RE2}),
we may pick up some simpler functional  equations such as 
\bea
\frac{\hat{z}2}{\hat{z}1}&=&
\frac{w_2(v,u)\tilde{\rho}_1(v,-u)\bar{z}2+w_3(v,u)\tilde{\rho}_2(v,-u)\bar{z}1}{w_3(-u,-v)\tilde{\rho}_2(u,-v)e^{-4\eta}\bar{z}2+w_2(-u,-v)\tilde{\rho}_1(u,-v)\bar{z}1},\label{ekp1}\\
\frac{\hat{z}3}{\hat{z}1}&=&
\frac{w_2(v,u)\tilde{\rho}_1(v,-u)e^{2\eta}\bar{z}3+w_3(v,u)\tilde{\rho}_2(v,-u)\bar{z}1}{w_3(-u
,-v)\tilde{\rho}_2(u,-v)\bar{z}3+w_2(-u,-v)\tilde{\rho}_1(u,-v)e^{2\eta}\bar{z}1},\\
\frac{\hat{z}5}{\hat{z}1}&=&
\frac{w_2(v,u)\tilde{\rho}_1(v,-u)\bar{z}5+w_3(v,u)\tilde{\rho}_2(v,-u)e^{-4\eta}\bar{z}1}{w_3(-
u,-v)\tilde{\rho}_2(u,-v)\bar{z}5+w_2(-u,-v)\tilde{\rho}_1(u,-v)\bar{z}1},\\
\frac{\hat{z}8}{\hat{z}6}&=&
\frac{w_{47}(-u,-v)\tilde{\rho}_3(u,-v)e^{2\eta}\bar{z}6+w_{46}(-u,-v)\tilde{\rho}_4(u,-v)\bar{z
}8}{w_{46}(v,u)\tilde{\rho}_4(v,-u)\bar{z}6+w_{47}(v,u)\tilde{\rho}_{3}(v,-u)e^{2\eta}\bar{z}8},
\\
\frac{\hat{z}6}{\hat{z}2}&=&
\frac{w_{19}(-u,-v)\tilde{\rho}_6(u,-v)e^{2\eta}\bar{z}2+w_{20}(-u,-v)\tilde{\rho}_5(u,-v)e^{4\e
ta}\bar{z}6}{w_{19}(v,u)\tilde{\rho}_6(v,-u)e^{2\eta}\bar{z}6+w_{20}(v,u)\tilde{\rho}_{5}(v,-u)\
bar{z}2},\\
\frac{\hat{z}4}{\hat{z}2}&=&
\frac{w_{20}(-u,-v)\tilde{\rho}_5(u,-v)\bar{z}2+w_{19}(-u,-v)\tilde{\rho}_7(u,-v)\bar{z}4}{w_{19
}(v,u)\tilde{\rho}_7(v,-u)\bar{z}2+w_{20}(v,u)\tilde{\rho}_{5}(v,-u)\bar{z}4},\\
\frac{\hat{z}7}{\hat{z}3}&=&
\frac{w_{19}(-u,-v)\tilde{\rho}_7(u,-v)e^{2\eta}\bar{z}7+w_{20}(-u,-v)\tilde{\rho}_5(u,-v)\bar{z
}3}{w_{19}(v,u)\tilde{\rho}_7(v,-u)e^{2\eta}\bar{z}3+w_{20}(v,u)\tilde{\rho}_{5}(v,-u)e^{4\eta}\
bar{z}7}.\label{ekp7}
\eea
Before further going on, we first need to work out the matrices $R_{21}^{{\rm t}_1{\rm
t}_2}(u,v)$, $R_{12}^{{\rm t}_1{\rm t}_2}(u,v)$,
$\tilde{R}_{12}(u,v)$ and $\tilde{R}_{21}(u,v)$ according to
their definitions in section 2. For our convenience, we prefer to present below some entries of 
$\tilde{R}_{12}$ involving in above equations as
\bea
\tilde{\rho}_1(u,v)& =& 
-\frac{(1+e^{6\eta}uv)(1+e^{4\eta}uv)^2(1+e^{2\eta}uv)(1+uv)}{(u-v)^2(e^{2\eta}u-v)(e^{2\eta}v-u
)(e^{4\eta}u-v)e^{4\eta}},\\
\tilde{\rho}_2(u,v)& =& 
\frac{(1+e^{6\eta}uv)(1+e^{4\eta}uv)^2(1+e^{2\eta}uv)(1+uv)\sqrt{(1+e^{4\eta}u^2)(1+e^{4\eta}v^2
)}}{(u-v)^2(e^{2\eta}u-v)(e^{2\eta}v-u)(e^{4\eta}u-v)(e^{4\eta}v-u)e^{2\eta}},\\
\tilde{\rho}_3(u,v)& =& 
\frac{(1+e^{4\eta}uv)(1+e^{2\eta}uv)(uv+e^{2\eta})(1+uv)^2\sqrt{(1+u^2)(1+v^2)}}{(u-v)^2(e^{2\eta}u-v)(e^{2\eta}v-u)(e^{4\eta}u-v)(e^{4\eta}v-u)},\\
\tilde{\rho}_4(u,v)& =& 
\frac{(1+e^{4\eta}uv)(1+e^{2\eta}uv)(uv+e^{2\eta})(1+uv)^2}{(u-v)^2(e^{2\eta}u-v)(e^{2\eta}v-u)(
e^{4\eta}v-u)},\\
\tilde{\rho}_5(u,v)& =& 
\frac{(1+e^{4\eta}uv)^2(1+e^{2\eta}uv)(1+uv)^2\sqrt{(1+e^{2\eta}u^2)(1+e^{2\eta}v^2)}}{(u-v)^2(e
^{2\eta}u-v)(e^{2\eta}v-u)(e^{4\eta}u-v)(e^{4\eta}v-u)},\\
\tilde{\rho}_6(u,v)& =&- 
\frac{(1+e^{4\eta}uv)^2(1+e^{2\eta}uv)(1+uv)^2}{(u-v)^2(e^{2\eta}u-v)(e^{2\eta}v-u)(e^{4\eta}u-v
)e^{\eta}},\\
\tilde{\rho}_7(u,v)& 
=&\frac{(1+e^{4\eta}uv)^2(1+e^{2\eta}uv)(1+uv)^2}{(u-v)^2(e^{2\eta}u-v)(e^{2\eta}v-u)(e^{4\eta}v
-u)e^{\eta}}.
\eea
Then by analying the structure of the Eqs. (\ref{ekp1})-(\ref{ekp7}), 
we can get the following relations: 
\bea
&&\frac{z2_+(u)}{z1_+(u)}=\frac{c_1u+e^{4\eta}c_8}{c_1u-c_8},~~~~~
\frac{z3_+(u)}{z1_+(u)}=\frac{c_2u+e^{2\eta}c_9}{e^{2\eta}c_2u-c_9},\\
& &
\frac{z5_+(u)}{z1_+(u)}=\frac{c_3u+c_{10}}{e^{4\eta}c_3u-c_{10}},~~~~~
\frac{z4_+(u)}{z2_+(u)}=\frac{c_6u+e^{2\eta}c_{11}}{e^{2\eta}c_6u-c_{11}},\\
& &
\frac{z6_+(u)}{z2_+(u)}=\frac{c_5u+c_{12}}{e^{4\eta}c_5u-c_{12}},~~~~~
\frac{z7_+(u)}{z3_+(u)}=\frac{c_7u+c_{13}}{e^{4\eta}c_7u-c_{13}},\\ &
&
\frac{z8_+(u)}{z6_+(u)}=\frac{c_4u+e^{2\eta}c_{14}}{e^{2\eta}c_4u-c_{14}}.
\eea Running second RE (\ref{RE2}) with above relations again and
again, the solutions (\ref{Kp}) would be fixed definitely.


\begin{thebibliography}{99}


\bibitem{BM86} J. B. Bednorz and K. A. M\"{u}ller, Z. Phys. {\bf B 64} (1986) 189.
\bibitem{[1]} V.E. Korepin and F.H.L. Essler: {\it Exactly
    Solvable Models of Strongly Correlated Electrons}, World
  Scientific, Singapore(1994)
\bibitem{[2]} P. Schlottmann, Phys. Rev. Lett.{\bf 68} (1992) 1916; S.
  Sarkar, J. Phys. {\bf A}: Math. Gen. {\bf 23} (1990) L409; P. A. Bares, G. Blatter and
M. Ogata, Phys. Rev. Lett. {\bf 73} (1991) 11340; I.N. Karnaukhov, Phys.
  Rev. Lett. {\bf73} (1994) 11340
\bibitem{[3]} F.H.L. Essler, V.E. Korepin and K.
  Schoutens, Phys. Rev. Lett.  {\bf 68} (1992) 2960; F.H.L. Essler and
  V.E. Korepin, Phys. Rev.   {\bf B 46} (1992) 9147
\bibitem{[4]} A.J.
  Bracken, M.D. Gould, J.R. Links and Y.Z. Zhang, Phys. Rev.  Lett.
    {\bf 74} (1995) 2768
\bibitem{LW} 
E.H. Lieb and F.Y. Wu, Phys. Rev. Lett.  {\bf 20} (1968) 1445
\bibitem{azr} P.W.~Anderson, Science {\bf 235} (1987) 1196; \newline
F.C.~Zhang and T.M.~Rice, Phys. Rev. {\bf B37} (1988) 3759
\bibitem{tj} B. Sutherland, Phys. Rev.  {\bf B 12} (1975) 3795;\newline 
P. Schlottmann, Phys. Rev.  {\bf B 36} (1987) 5177;\newline
P. Wiegmann, Phys. Rev. Lett.  {\bf 60} (1988) 821;\newline
F.H.L.~Essler and V.E.~Korepin, Phys. Rev. {\bf B46}
(1992) 9147;\newline
A.~Foerster and M.~Karowski, Phys. Rev.
{\bf B 46} (1992) 9234; Nucl. Phys. {\bf B396} (1993) 611
\bibitem{Bar1}R.Z. Bariev, J. Phys.  {\bf A}: Math. Gen.  {\bf 24} (1991) L549;  {\bf A}: Math. 
Gen.  {\bf 24} (1991) L919.
\bibitem{Bar2}R.Z. Bariev, A. Kl\"umper, A. Schadschneider and J. 
Zittartz, J. Phys. {\bf A}: Math. Gen.  {\bf 26} (1993) 4663;  {\bf A}:
Math. Gen.   {\bf 26} (1993) 1249
\bibitem{qism1} E. K. Sklyanin and L.D. Faddeev, Sov. Phys. Dokl. {\bf 23} (1978) 902;\newline
E.K. Sklyanin, J. Sov. Math. {\bf 19} (1982) 1546.
\bibitem{qism3}P.P. Kulish and E.K. Sklyanin, Lecture Notes in Physics {\bf 151} (1982)
61, Berlin: Springer-verlag.
\bibitem{qism4} V.E. Korepin, N.M. Bogoliubov and A.G. Izergin, 
{\it  Quantum inverse Scattering Method and Correlation Function},
  Cambridge University Press 1993
\bibitem{op1}F.C. Alcaraz, M.N. Barber, M.T. Batchelor, R.J. Baxter and G.R.W. Quispel,
J. Phys. {\bf A }: Math. Gen.  {\bf 20} (1987) 6397;\newline
 I.V. Cherednik, Theor. Math. Phys.  {\bf 61} (1984) 911.
\bibitem{op2} E.K.Sklyanin, J. Phys. {\bf A }: Math. Gen.  {\bf 21} (1988) 2375.
\bibitem{op3} L. Mezincescu and R.I. Nepomechi, J. Phys.   {\bf A }:Math. Gen.   {\bf 24} (1991) 
L17;\newline 
Int. J. Mod. Phys. {\bf  A7} (1991)5231 ; Int. J. Mod. Phys.  {\bf A7}
(1992)5657.
\bibitem{op4}H.Q. Zhou, X.Y. Ge, J.R. Links and M.D. Gould, Nucl. Phys. 
 {\bf B 546} (1999) 779
\bibitem{Frah}G. Bed\"{u}rftig and H. Frahm, Phsica  {\bf E 4} (1999) 246;
J. Phys.  {\bf A}: Math. Gen.  {\bf 32} (1999) 4585;\newline
G. Bed\"{u}rftig and H. Frahm, J. Phys. {\bf A}: Math. Gen.  {\bf 30}, (1997) 4139;\newline
G. Bed\"{u}rftig , B. Brendel, H. Frahm and R.M. Noack, Phys. Rev. {\bf B 58} (1998) 10225
\bibitem{Asak}H. Asakawa and M. Suzuki, Physica  {\bf A 236} (1997) 376; 
H. Asakawa,  Physica  {\bf A 256} (1998) 229.
\bibitem{Wad1}
H. Asakawa and M. Suzuki, J. Phys.  {\bf A}: Math. Gen.   {\bf 29} (1996) 225;\newline
M. Shiroishi and M. Wadati, J. Phys. Soc. Jpn,  {\bf 66} (1997) 1
\bibitem{be1} S. Skorik and A. Kapustin,  J. Phys.  {\bf A}: Math. Gen. {\bf 29} (1996) 
1629;\newline
 S. Skorik and H. Saleur, J. Phys.  {\bf A}: Math. Gen.  {\bf 28} (1995) 6605. 
\bibitem{be2} A.A. Zvyagin, Phys. Rev.  {\bf B60} (1999) 15266 ; \newline
A.A. Zvyagin and 
P. Schlottmann, Phys. Rev.  {\bf B56 } (1997) 300;\newline
 A.A. Zvyagin and H. Johannesson, Phys. Rev. Lett. {\bf 81} (1998) 2751.
\bibitem{be3} 
A. Foerster and M. Karowski, Nucl. Phys.{\bf B 408} (1993) 512;\newline
 A.J. Bracken, X.Y. Ge, Y.Z. Zhang and H.Q. Zhou, 
   Nucl. Phys. {\bf B 516} (1998) 588;\newline
X.-W. Guan, J. Phys.  {\bf A}: Math. Gen.  {\bf 33} (2000) 5391
\bibitem{op5} 
H.Q. Zhou, Phys. Rev.  {\bf B 54 }(1996) 41; \newline
X.-W. Guan, M.-S. Wang and S.-D. Yang, Nucl. Phys. {\bf B 485} (1997) 685;\newline
M. Shiroishi and M. Wadati, J. Phys. Soc. Jpn.  {\bf 66} (1997) 2288;\newline
T. Deguchi, R.Yue and K.  Kusakabe, J. Phys.  {\bf A}: Math. Gen.  {\bf 31} (1998) 7315
\bibitem{op6}Z.-N Hu and F.-C. Pu, Nucl. Phys.  {\bf B 546} (1999) 691;\newline
H. Fan, M. Wadati and X.-M. Wang,
Phys. Rev.  {\bf B 61} (2000) 3450;
A.Lima-Santos, Nucl. Phys. {\bf B 558} (1999) 637
\bibitem{bariev}H.-Q. Zhou, Phys. Rev. {\bf B 53} (1996) 5098;\newline
 A. Foerster, X.-W. Guan, J. Links, I. Roditi and H.-Q. Zhou,
 Nucl. Phys. {\bf B 596} (2001) 525
\bibitem{Zhou97} H.-Q. Zhou and D.-M. Tong,
    Phys. Lett. {\bf  A 232} (1997) 377
\bibitem{zhou3}A.J. Bracken, X.-Y. Ge, Y.-Z. Zhang and H.-Q. Zhou,
Nucl. Phys. {\bf B 516} (1998) 603
\bibitem{deV93} H.J. de Vega and A. Gonz\'alez-Ruiz, J. Phys. A:
   Math. Gen.  {\bf 26} (1993) L519; Mod. Phys. Lett.{\bf  A 9} (1994)
   2207; \newline     
 J. Abad and M. Rios, Phys. Lett. {\bf  B 352} (1995) 92.

\end{thebibliography}
\end{document}